\documentstyle[12pt,epsf]{article} \hoffset -0.5in \textwidth 6.00in \textheight
8.6in  \parskip 7pt \openup1\jot \parindent=0.5in
\newskip\humongous \humongous=0pt plus 1000pt minus 1000pt
\def\caja{\mathsurround=0pt}
\def\eqalign#1{\,\vcenter{\openup1\jot \caja
        \ialign{\strut \hfil$\displaystyle{##}$&$
        \displaystyle{{}##}$\hfil\crcr#1\crcr}}\,}
\newif\ifdtup

\def\eqright #1\cr{\noalign{\hfill$\displaystyle{{}#1}$}}
\def\eqleft #1\cr{\noalign{\noindent$\displaystyle{{}#1}$\hfill}}

\def\oldreffmt#1{\rlap{[#1]} \hbox to 2\parindent{}}

\def\figfmt#1{\rlap{Figure {#1}} \hbox to 1in{}}

\def\VEV#1{\left\langle #1\right\rangle}

\def\sectioneq{\def\theequation{\thesection.\arabic{equation}}{\let
\holdsection=\section\def\section{\setcounter{equation}{0}\holdsection}}}%

\newcounter{holdequation}

\def\begineq #1\endeq{$$ \refstepcounter{equation}\eqalign{#1}\eqno
	(\theequation) $$}
\def\contlimit{\,{\hbox{$\longrightarrow$}\kern-1.8em\lower1ex
\hbox{${\scriptstyle (a\rightarrow0)}$}}\,}
\def\centeron#1#2{{\setbox0=\hbox{#1}\setbox1=\hbox{#2}\ifdim
\wd1>\wd0\kern.5\wd1\kern-.5\wd0\fi
\copy0\kern-.5\wd0\kern-.5\wd1\copy1\ifdim\wd0>\wd1
\kern.5\wd0\kern-.5\wd1\fi}}
\def\centerover#1#2{\centeron{#1}{\setbox0=\hbox{#1}\setbox
1=\hbox{#2}\raise\ht0\hbox{\raise\dp1\hbox{\copy1}}}}
\def\centerunder#1#2{\centeron{#1}{\setbox0=\hbox{#1}\setbox
1=\hbox{#2}\lower\dp0\hbox{\lower\ht1\hbox{\copy1}}}}
\def\lsim{\;\centeron{\raise.35ex\hbox{$<$}}{\lower.65ex\hbox
{$\sim$}}\;}
\def\gsim{\;\centeron{\raise.35ex\hbox{$>$}}{\lower.65ex\hbox
{$\sim$}}\;}

\def\super#1{\ifmmode \hbox{\textsuper{#1}}\else\textsuper{#1}\fi}
\def\textsuper#1{\newcount\holdspacefactor\holdspacefactor=\spacefactor
$^{#1}$\spacefactor=\holdspacefactor}

\def\getcite#1,{\advance\citenumber by1
\def\getcitearg{#1}\def\lastarg{@}
\ifnum\citenumber=1
\ref{#1}\let\next=\getcite\else\ifx\getcitearg\lastarg\let\next=\relax
\else ,\ref{#1}\let\next=\getcite\fi\fi\next}

\def\pom{{\rm P\kern -0.53em\llap I\,}}
\def\spom{{\rm P\kern -0.36em\llap \small I\,}}
\def\sspom{{\rm P\kern -0.33em\llap \footnotesize I\,}}

\relax

\def\begineq #1\endeq{$$ \refstepcounter{equation}\eqalign{#1}\eqno
	(\theequation) $$}
\def\contlimit{\,{\hbox{$\longrightarrow$}\kern-1.8em\lower1ex
\hbox{${\scriptstyle (a\rightarrow0)}$}}\,}
\def\centeron#1#2{{\setbox0=\hbox{#1}\setbox1=\hbox{#2}\ifdim
\wd1>\wd0\kern.5\wd1\kern-.5\wd0\fi
\copy0\kern-.5\wd0\kern-.5\wd1\copy1\ifdim\wd0>\wd1
\kern.5\wd0\kern-.5\wd1\fi}}
\def\centerover#1#2{\centeron{#1}{\setbox0=\hbox{#1}\setbox
1=\hbox{#2}\raise\ht0\hbox{\raise\dp1\hbox{\copy1}}}}
\def\centerunder#1#2{\centeron{#1}{\setbox0=\hbox{#1}\setbox
1=\hbox{#2}\lower\dp0\hbox{\lower\ht1\hbox{\copy1}}}}
\def\lsim{\;\centeron{\raise.35ex\hbox{$<$}}{\lower.65ex\hbox
{$\sim$}}\;}
\def\gsim{\;\centeron{\raise.35ex\hbox{$>$}}{\lower.65ex\hbox
{$\sim$}}\;}

\def\super#1{\ifmmode \hbox{\textsuper{#1}}\else\textsuper{#1}\fi}
\def\textsuper#1{\newcount\holdspacefactor\holdspacefactor=\spacefactor
$^{#1}$\spacefactor=\holdspacefactor}

\def\getcite#1,{\advance\citenumber by1
\ifnum\citenumber=1
\ref{#1}\let\next=\getcite\else\ifx#1@\let\next=\relax
\else ,\ref{#1}\let\next=\getcite\fi\fi\next}

\def\upon #1/#2 {{\textstyle{#1\over #2}}}
\relax

\def\mainhead#1{\setcounter{equation}{0}\addtocounter{section}{1}
  \vbox{\begin{center}\large\bf #1\end{center}}\nobreak\par}
\sectioneq

\def\til#1{\centeron{\hbox{$#1$}}{\lower 2ex\hbox{$\char'176$}}}
\def\tild#1{\centeron{\hbox{$\,#1$}}{\lower 2.5ex\hbox{$\char'176$}}}
\def\sumtil{\centeron{\hbox{$\displaystyle\sum$}}{\lower
-1.5ex\hbox{$\widetilde{\phantom{xx}}$}}}

\def\pom{{\rm P\kern -0.53em\llap I\,}}
\def\spom{{\rm P\kern -0.36em\llap \small I\,}}
\def\sspom{{\rm P\kern -0.33em\llap \footnotesize I\,}}

\newcommand{\bit}{\begin{itemize}}
\newcommand{\eit}{\end{itemize}}

\newcommand{\beq}{\begin{equation}}
\newcommand{\eeq}{\end{equation}}
\newcommand{\beqa}{\begin{eqnarray}}
\newcommand{\eeqa}{\end{eqnarray}}

\begin{document} 

\rightline{\vbox{\halign{&#\hfil\cr
&ANL-HEP-CP-97-47 \cr
&\today\cr}}} 
\vspace{1.25in} 

\begin{center} 
 
{\large\bf 
Excess Cross-Sections at the Electroweak Scale in the Sextet Quark
``Standard Model'' }\footnote{Work 
supported by the U.S.
Department of Energy, Division of High Energy Physics, \newline Contracts
W-31-109-ENG-38 and DEFG05-86-ER-40272} 
\medskip

Alan. R. White\footnote{arw@hep.anl.gov }

\vskip 0.6cm

\centerline{High Energy Physics Division}
\centerline{Argonne National Laboratory}
\centerline{9700 South Cass, Il 60439, USA.}
\vspace{0.5cm}

\end{center}

\begin{abstract} 

If dynamical electroweak symmetry breaking is due to a flavor doublet of
color sextet quarks, enhanced electroweak scale QCD instanton interactions 
may produce a large top mass, raise the $\eta_6$ axion mass, and also
explain the excesses in the DIS cross-section at HERA and jet cross-sections
at the Tevatron. 

\end{abstract} 

\vspace{1.5in}
\begin{center}
Presented at the 5th International Workshop on Deep Inelastic 
Scattering and QCD, Chicago, Illinois, USA, April 14-18, 1997
\end{center}

\newpage

\mainhead{1. INTRODUCTION} 
 
It is possible that the large $x$ and $Q^2$ events seen at
HERA are simply an excess in the cross-section compared to the standard model
prediction. Since the scales are similar, the same physics 
could also be responsible for the large $E_T$ excess in the jet
cross-section observed by CDF at the Tevatron. If this is the
case, these phenomena could be a crucial pointer towards improvement of the
Standard Model. In this talk\footnote{See hep-ph/9704248
for a fuller account, together with 
a full set of references.} I will describe the ``sextet quark model''
(SQM) obtained by replacing the Higgs sector of the Standard Model with a
flavor doublet of color sextet quarks. In this model both QCD and the
electroweak interaction are modified above the electroweak scale, and a
number of phenomena are inter-related. In particular enhanced instanton
interactions are important for electroweak-scale CP violation, the axion
mass, the top mass and new, {\it chirality violating}, light quark 
interactions. 

\mainhead{2. ELECTROWEAK SYMMETRY BREAKING }

We add to the Standard Model (with no scalar Higgs sector), a massless 
flavor doublet $Q\equiv(U,D)$ of color sextet quarks with the {\em usual
quark quantum numbers}, except that the role of quarks and antiquarks is
interchanged. (To cancel the $SU(2)\otimes U(1)$ anomaly other fermions with 
electroweak quantum numbers must also be added.) 
We expect that the axial part of the $U(2)\otimes U(2)$ chiral flavor symmetry 
breaks spontaneously and produces four massless pseudoscalar
mesons - $\pi^+_6,\;\pi^-_6,\;\pi^0_6$
and $\eta_6$. Since the pseudoscalars couple longitudinally 
to the sextet axial currents, the usual technicolor mechanism leads to
the $\pi^+_6,\;\pi^-_6$ and $\pi^0_6$ becoming, respectively, the third
components of the $W^+,\;W^-$ and $Z^0$, with $M_W\sim g\;F_{\pi_6}$, 
where $F_{\pi_6}$ is the sextet QCD chiral scale. The ``Casimir Scaling'' rule
$C_6\alpha_s(F^2_{\pi_6})~\sim~C_3\alpha_s(F^2_{\pi})
$ (with $C_6/C_3 ~=~5/2~$) 
is clearly consistent with $F_{\pi_6}\sim 250$ GeV ! 
Asymptotic freedom allows only one sextet doublet and so $
\rho=~M^2_W/M^2_Zcos^2\theta_W~=~1
$, 
as required by experiment. 

\mainhead {3. THE $\eta_6$ AND CP VIOLATION}

The $\eta_6$ couples to the QCD color anomaly, and is {\it an axion}.
In conventional QCD, instanton interactions give only a very small indirect
contribution to the 
axion mass. As we discuss below, within the SQM instanton interactions are
strongly enhanced above the electroweak scale. 
Adding instanton and anti-instanton interactions produces factors of 
$\cos [\tilde{\theta}+ \VEV{\eta_6}]$ so that the axion potential generated
naturally retains the $CP$-conserving 
minimum at $~\tilde{\theta}+\VEV{\eta_6} =0$ while simultaneously producing
a large contribution to the $\eta_6$ mass. 

Above the
electroweak scale, we can not write a lagrangian 
involving both the $\eta_6$ and the gluon field to describe general sextet 
quark 
interactions. We must use the full
$QCD$ lagrangian, which clearly has no axion. Hence QCD interactions above
the electroweak scale will naturally be {\it Strong $CP$-violating}. 
The triplet quark mesons will contain a 
small admixture of sextet quark states which will provide 
$CP$ violating interactions. Therefore {\it electroweak scale $CP$-violation
may actually be ``Strong $CP$-violation'' within the SQM.}

\mainhead {4. THE QCD $\beta$-FUNCTION AND WALKING COLOR}

Adding two sextet flavors, the QCD $\beta$-function
has an infra-red fixed point at $\alpha_s \approx  1 /  34$.
Between the ultra-violet and infra-red fixed points $\beta(\alpha_s)$
remains very small ($ <10^{-6}$). As a result, 
$\alpha_s$ will dramatically 
stop it's conventional evolution at some scale and will then evolve 
extremely slowly. (It is possible that $\alpha_s$ 
can be defined via jet cross-sections in such a way that the 
evolution stops just above the electroweak scale, and so effectively 
explains the CDF excess.)
When $\beta(\alpha_s)$ is approximated as a small constant, 
the linearized Dyson-Schwinger equation 
has a solution for the dynamical (sextet) mass of the form
$ \Sigma (p) \sim ~\mu^2~(p)^{-1}$.
When this behavior is inserted into the perturbative
formula for the high-momentum component of the sextet condensate
$\VEV{Q\bar{Q}}$, there is a strong enhancement. 
There is no corresponding enhancement of the chiral constant. 

\mainhead {5. ENHANCED INSTANTON INTERACTIONS}

Because of the infra-red fixed-point, QCD instanton interactions have no
infra-red divergences in the massless SQM. In the (``physical'')  massive SQM 
these interactions provide all of the non-perturbative physics down to the 
electroweak scale. The extremely slow evolution of $\alpha_s$ implies 
that there is almost no short-distance cut-off for such interactions.

The large sextet Casimir leads to a surprizingly high-order one 
instanton interaction. 
The one instanton zero modes produce an interaction which
involves one quark/antiquark pair of each triplet
flavor and five pairs of each sextet flavor.
If we close-up sextet lines pairwise
with the sextet condensate, we obtain the usual QCD triplet interaction 
- enhanced by a factor of $\VEV{Q\bar{Q}}^{10}$. The resulting (electroweak 
scale) contribution to the triplet quark mass matrix is
$ \Sigma_3~\sim ~det~m^0_3 ~[m^0_3]^{-1} 
~\equiv ~\prod_{j \neq i}~ m_j^0 $, ~effectively inverting the ``bare'' mass 
matrix $m^0_3$. If we make the (extremely oversimplifying) assumption that
the single instanton interaction represents the dynamical effects of
instantons between an upper (cut-off) scale and the electroweak scale, the
electroweak scale top 
quark mass can be explained as a consequence of an 
anomalously small bare mass. In addition, above the electroweak scale, {\it the
light quark dynamical masses will become comparable to that of the top
quark mass.} 

There are also four-quark
interactions ($V{qq}$) involving pairs of quarks with distinct
flavors. The largest interactions will involve the top quark, if indeed it 
has the smallest bare mass. Chirality non-conserving light quark couplings to
a gauge field are obtained from ``higher-order'' interactions 
of the form
\begin{center}

\leavevmode
\epsfxsize=4in
\epsffile{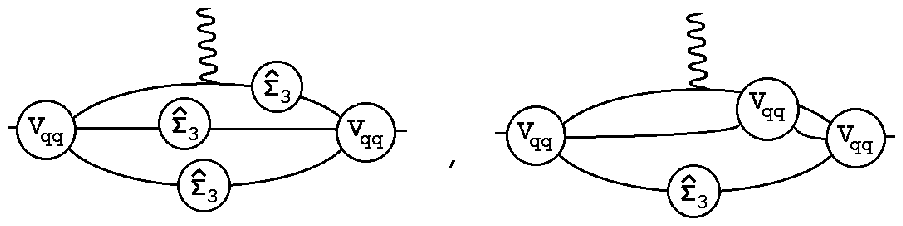}
\end{center}
For such interactions 
chirality-violating and chirality-conserving amplitudes  should be
comparable in order of magnitude.

Since the bare triplet quark masses badly break the relevant U(1) symmetry,
a large $\eta_6$ mass arises straightforwardly from sextet instanton 
interactions. A very crude estimate gives a mass close to, or not too far
below, the electroweak scale. 
The $\eta_6$ will be produced in association with 
the $W$ or the $Z$ 
and could decay 
predominantly into  $b\bar{b}$ states. 
(This could lead to confusion 
with the standard model Higgs experimently.) 

\mainhead {6.  EXCESS CROSS-SECTIONS}

At HERA the excess cross-section appears for
$ x ~\centerunder{\raisebox{1mm}{$\scriptscriptstyle >$}}
{$\scriptscriptstyle \sim$}~0.5 ~\leftrightarrow ~xP~
\centerunder{\raisebox{1mm}{$\scriptscriptstyle >$}}
{$\scriptscriptstyle \sim$}~400~GeV$ and
$Q~\centerunder{\raisebox{1mm}{$\scriptscriptstyle >$}}
{$\scriptscriptstyle \sim$}~150~GeV$    
and so is kinematically just where we expect the new instanton vertices 
to contribute. Similarly, the CDF jet excess is for $E_T 
\centerunder{\raisebox{1mm}{$\scriptscriptstyle >$}}
{$\scriptscriptstyle \sim$}~200~GeV$. At lowest order in the instanton 
interaction, only the light quark dynamical mass terms contribute in both cases.
Although we expect higher-order and multi-instanton interactions to be 
essential quantitatively, the dynamical masses give some qualitative idea of 
the properties to be expected. In particular dynamical masses give -
\newline $~~~~~\raisebox{-1mm}{*}~$ additional helicity-flip quark final states 
\newline $~~~~~\raisebox{-1mm}{*}~$ quark jet cross-sections increase with 
the jet mass
\newline $~~~~~\raisebox{-1mm}{*}~$  jet angular distributions are unchanged 
\newline $~~~~~\raisebox{-1mm}{*}~$  quark parton distributions are enhanced 
at large $x$
\newline $~~~~~\raisebox{-1mm}{*}~$  the presence of gluon initial and final
states at the Tevatron implies 
\newline $~~~~~~~~$ effects of the new interactions will be less
dramatic than at HERA.

\end{document}